\documentclass[12pt,a4paper]{article}

\usepackage{amsmath}
\usepackage{graphicx}
\usepackage{times}
\usepackage{cite}
\usepackage{ulem}
\begin{document}
\begin{titlepage}
\title{{Comments on oscillations  in  elastic scattering of hadrons}}
\author{ S.M. Troshin, N.E. Tyurin\\[1ex]
\small  \it NRC ``Kurchatov Institute''--IHEP\\
\small  \it Protvino, 142281, Russian Federation,\\
\small Sergey.Troshin@ihep.ru
}
\normalsize
\date{}
\maketitle

\begin{abstract}
	The observed absence  of the small-$t$ oscillations in differential cross--section of the elastic scattering is considered as a consequence of 
the reflective scattering mode 	appearance  at the highest energy of the LHC $\sqrt{s}=13$ TeV.
\end{abstract}
\end{titlepage}
\setcounter{page}{2}

 The recent analysis of the  LHC measurements at the  energy of $\sqrt{s}=13$ TeV indicated an absence of oscillations in small-$t$ differential cross--section of $pp$ elastic scattering \cite{graf}. Existence of such oscillations at lower energies has been advocated in \cite{sel0}. Their possible origin  discussed e.g. in \cite{sel}  is often associated with the phenomenon known under abreviation of ``AKM--scaling'' by the names of authors (Auberson, Kinoshita and Martin)  \cite{akm}. The AKM scaling and related oscillations  have been obtained in the limit of $s\to\infty$, and their discussion  may be considered  as a rather irrelevant in view of the data at $\sqrt{s}=13$ TeV. However, the experimental result itself seems very interesting and it additionally confirms  reliability of the ATLAS and TOTEM extrapolations of $d\sigma/dt$ to the point $-t=0$ \cite{atl,ttm}. 
In fact,  the AKM oscillations  of the differential cross--section are more relevant to the dip region of $d\sigma/dt$.  It should be noted that  a presence of the  coherent structures (oscillations) in the differential cross-sections of hadron scattering is considered  even in the region of a fixed--angle scattering \cite{hen}.
 
 Thus, we take for granted an existence of small-$t$ oscilations at lower energies, regardless of their interpretation, as well as their dissapearance at $\sqrt{s}=13$ TeV. 
 
 Our intension is to present    qualitative considerations on  possible physical origin of such energy behavior of  the fine--grained structure in the diffraction peak of the elastic scattering differential cross--section.
 
 The most evident explanation for the  oscillations vanishing  is related to  evolution of the helicity--changing amplitudes with the energy growth.
 These amplitudes start to contribute at $-t\neq 0$ and it  could lead to oscillations in the differential cross--section of elastic scattering.  
  Due to degradation  of the spin role with the energy increase, the helicity--changing amplitudes become irrelevant, and the oscillations dissapear.
 
 However,  the role of  spin  is often neglected from the very begining. In the case, one can justify the observed phenomenon of oscillations vanishing by  emergence of  reflective scattering mode \cite{refl}. 
  
 The reflective scattering mode results in  formation of a peripheral impact parameter dependence of the inelastic overlap function.  Emergent peripheral  form of the inelastic overlap is associated with self--dumping of the inelastic channels contribution to unitarity at the LHC energies. Such self-dumping can arise due to randomization of the phases of  the multiparticle production amplitudes.  This  randomization in its turn can be considered as a consequence of  formation of a color--conducting collective hadronic matter under the central  hadron collisions  (with high multiplicities) since   subsequent stochastic decay of this state and hadronization   occur \cite{jpg}.  
 
 Thus, we propose to associate oscillations and their disappearance with the unitarity and its saturation at $s\to\infty$. Unitarity and the shadow scattering  mode have been used in \cite{pred} for  explanation of the diffraction peak formation in the elastic scattering wherein randomization of the multiparticle production amplitudes phases was the main assumption. Those phases play a nontrivial and important role for obtaining  correct value of the slope parameter\cite{koba}. Unitarity also generates  growth  of the slope parameter  with energy \cite{gener}. We will take into account the reflective scattering mode emergence being a consequence of the unitarity  saturation at asymptotics.
 
 We write unitarity equation for the elastic scattering amplitude $F(s,t)$  in the form
 \begin{equation} \label{unit}
 	\mbox{Im} F(s,t)=H_{el}(s,t)+H_{inel}(s,t),
 \end{equation}
where $H_{el}(s,t)$ is the two--particle intermidiate state contribution  and  $H_{inel}(s,t)$ is the sum of the  contributions from the multi--particle intermidiate states. For the forward scattering when $-t=0$ the  Eq. (\ref{unit}) turns into
 \begin{equation} \label{unit0}
 	\sigma_{tot}(s)=\sigma_{el}(s)+\sigma_{inel}(s),
 \end{equation}
where $\sigma_i(s)$ for $i=tot,el,inel$ are the respective cross--sections. 
High--energy elastic scattering amplitude is a predominantly imaginary and is given by the sum of the two contributions $H_{el}(s,t)$ and  $H_{inel}(s,t)$.

The reflective scattering mode corresponds to  dominance  of  the elastic overlap function $H_{el}(s,t)$. 
In the impact parameter representation the elastic and inelastic overlap functions have different profiles at high energies. The elastic overlap function preserves central profile. Contrary, the inelastic overlap function becomes peripheral.  Respectively, the term $H_{el}(s,t)$  in  Eq. (\ref{unit}) dominates at $s\to\infty $. As it was shown,   the reflective ability  of the interaction region \cite{prefa} comes into play at the energy value of $\sqrt{s}=13$ TeV.   We associate   vanishing the oscillations in $d\sigma/dt$ of the elastic pp-scattering with emergence of the reflective scattering mode. Respectively, the fine--grained structure of the diffraction cone remains to be a feature of the shadow or absorption scattering.

Now, we would like to comment on possible origin of the observed oscillations in $d\sigma/dt$ of the elastic $pp$-scattering at lower energies which correspond to absorptive or shadow scattering
\cite{trans}.

Driving force of elastic scattering in the absorptive/shadow mode is the contribution of  intermidiate multiparticle states into the unitarity equation represented by  inelastic overlap function $H_{inel}(s,t)$ in Eq. (\ref{unit}). Adopting notations of \cite{pred} this function can be written in the form
\begin{equation}\label{sum}
H_{inel}=\sum\int_nc_nF_{2n}F_{\tilde 2n}^*,
\end{equation}	
where $F_{2n}$ is the amplitude of $2\to n$ transition    and $n$ counts all admissable states at given energy. Number of such states increases with the collision energy. Functions $F_{2n}$ and $F_{\tilde 2n}$ have different arguments, they  differ in the initial states: one of the initial state is rotated by the scattering angle around the axis perpendicular to the scattering plane \cite{koba}. The coefficients $c_n$ are positive and account for the energy momentum conservation under $2\to n$ transition. The amplitudes and their phases should have strong dependence on the momenta of produced particles \cite{koba,krz} and such a behavior  could be a  source of the fine--grained structure (oscillations in the small--$t$ region) in the elastic  scattering at $-t\neq0$ under lower energies where the elastic scattering has  a shadow nature.

Appearance of  the reflective scattering mode is able to mask  the small--$t$  oscillations since the scattering amplitude becomes  progressively decoupled from dynamics of the multiparticle production processes. Such an evolution become evident under description in terms of the impact parameter dependent quantities \cite{trans}.

Finally, we would like to emphasize again: one should distinguish a fine--grained structure of the forward peak in the elastic scattering from the dips and bumps in $d\sigma/dt$. Those two structures are of a different physical origin. A coarse--grained structure can exist in  the both modes , while fine--grained structure should be associated with  the shadow scattering mode only.  

Thus, difficulties with quantitative estimations of fine--grained structure are evident: one needs to solve a formidable task of the multiparticles' amplitudes parameterization including their modules and phases. Additional problem  arises due to  quantum effects contribution from particles  entanglement \cite{zu}. Effects of quantum entanglement should also be taken into account \cite{khar} and those could also introduce particle correlations in the final state along with the fine--grained structure in the elastic scattering under the shadow scattering mode. 

Early approaches to  the multiparticle amplitude modeling have been described in \cite{mich,mich1,mich2, koba,lkh}. It is rather interesting to note that the uncorrelated jet model for hadron production proposed in \cite{ujm} has several common features with the later developed parton model (see for the recent accounting of the model \cite{khar}), especially, in respects of a presence of the unobservable  phases \cite{mich2}. The stochastic approach to shadow scattering has been considered in \cite{lkh} and  fluctuations of a number of particles in the intermidiate state of elastic scattering reaction can also be manifested as a fine--graned structure of differential cross--section according to  a representation of scattering  as a particle transfer through an anisotropic absorptive substance\cite{lkh}. The inelastic overlap function represents the summary impact of the inelastic channels onto elastic scattering in the shadow scattering mode.
The recent attempts  to apply the inelastic overlap function modelling for restoration of the elastic scattering amplitude at small values of $-t$ have been performed in \cite{rbb,univ}.

Leaving aside any particular model, we would argue  that the fine structure in $d\sigma/dt$ seems to be a feature associated with  shadow scattering   and its vanishing,  observed in the elastic $pp$-scattering at $\sqrt{s}=13$ TeV,  indicates  a start of the reflective scattering mode. Under  its emergence, the elastic scattering dynamics gradually decouples from multiparticle production dynamics.  The multiparticle production in its turn  does changes  its character too, it progressively becomes peripheral  \cite{int} and  loses  its driving ability for the elastic scattering  under the collision energy growth.  

What was said above is no way a  quantitative statement but a rather hypothetical conclusion. It is evident that the issues related to   the elastic scattering differential cross--section behavior deserve  further studies.

\end{document}